\documentclass[12pt]{article}
\usepackage{bbm}
\usepackage{epic}
\newcommand{\be}{\begin{equation}}
\newcommand{\ee}{\end{equation}}
\newcommand{\ba}{\begin{eqnarray}}
\newcommand{\ea}{\end{eqnarray}}
\newcommand{\no}{\nonumber\\}
\newcommand{\lesssim}{\:\mbox{\raisebox{-3pt}{$\stackrel%
{\displaystyle <}{\sim}$}}\:}

\begin{document}
\title{\normalsize \hfill UWThPh-2004-14 
\\[3mm] \LARGE
Lepton mixing angle $\theta_{13} = 0$ \\
with a horizontal symmetry $D_4$}
\author{Walter Grimus,$^a$\thanks{E-mail: walter.grimus@univie.ac.at}
\setcounter{footnote}{2}
Anjan S.\ Joshipura,$^b$\thanks{E-mail: anjan@prl.ernet.in} 
\hspace{5pt}
Satoru Kaneko,$^c$\thanks{E-mail: satoru@phys.ocha.ac.jp}
\hspace{5pt} \\
\setcounter{footnote}{5}
Lu\'\i s Lavoura,$^d$\thanks{E-mail: balio@cfif.ist.utl.pt} 
\hspace{5pt} and$\!$
\setcounter{footnote}{6}
Morimitsu Tanimoto$^e$\thanks{E-mail: tanimoto@muse.sc.niigata-u.ac.jp} 
\\*[3mm] 
\small $^a$ Institut f\"ur Theoretische Physik,
Universit\"at Wien \\[-1mm]
\small Boltzmanngasse 5, A--1090 Wien, Austria \\[1mm]
\small $^b$ Physical Research Laboratory \\[-1mm]
\small Ahmedabad 380009, India \\[1mm]
\small $^c$
Department of Physics,
Ochanomizu University \\[-1mm]
\small Tokyo 112-8610, Japan \\[1mm]
\small $^d$ Instituto Superior T\'ecnico,
Universidade T\'ecnica de Lisboa \\[-1mm]
\small P--1049-001 Lisboa, Portugal \\[1mm]
\small $^e$ Department of Physics, Niigata University \\[-1mm]
\small Ikarashi 2-8050, 950-2181 Niigata, Japan}

\date{9 July 2004}

\maketitle

\vspace{-6mm}

\begin{abstract}
We discuss a model for the lepton sector
based on the seesaw mechanism and on a $D_4$ family symmetry.
The model predicts the mixing angle $\theta_{13}$ to vanish.
The solar mixing angle $\theta_{12}$ is free---it will in general
be large if one does not invoke finetuning.
The model has an enlarged scalar sector with three Higgs doublets,
together with two real scalar gauge singlets $\chi_i$
($ i = 1, 2$)
which have vacuum expectation values
$\left\langle \chi_i \right\rangle_0$ at the seesaw scale. 
The atmospheric mixing angle $\theta_{23}$ is given
by $\tan \theta_{23} = \left\langle \chi_2 \right\rangle_0
\left/ \left\langle \chi_1 \right\rangle_0 \right.$,
and it is maximal if the Lagrangian is $D_4$-invariant;
but $D_4$ may be broken softly,
by a term of dimension two in the scalar potential,
and then $\left\langle \chi_2 \right\rangle_0$
becomes different from $\left\langle \chi_1 \right\rangle_0$.
Thus,
the strength of the soft $D_4$ breaking
controls the deviation of $\theta_{23}$ from $\pi / 4$. 
The model predicts a normal neutrino mass spectrum
($m_3 > m_2 > m_1$)
and allows successful leptogenesis if
$m_1 \sim 4 \times 10^{-3}\, \mathrm{eV}$;
these properties of the model are independent
of the presence and strength of the soft $D_4$ breaking.
\end{abstract}

\newpage

\section{Introduction} \label{intro}

The idea of neutrino oscillations~\cite{pontecorvo}
through neutrino masses and lepton mixing~\cite{MNS}
has proved successful and has led to continuous theoretical
and experimental efforts and progress---for a review see,
for instance,
\cite{maltoni}.
Still,
no favoured theory for explaining the observed patterns
of neutrino masses and lepton mixing has emerged yet;
in recent years,
myriads of textures and models for the lepton mass matrices
have been proposed---for a review see,
for instance,
\cite{altarelli}.
Only the smallness of neutrino masses
has found a favourite explanation in the seesaw mechanism~\cite{seesaw}.
In this paper we employ that mechanism,
and dismiss textures,
trying instead to trace some features of the lepton sector
to family \emph{symmetries}. 
In particular,
one may try and elucidate whether there is a connection
between the smallness of the mixing angle $\theta_{13}$
and the maximality of the atmospheric mixing angle $\theta_{23}$.
It turns out that there is no such connection:
\begin{itemize}
\item In the model of~\cite{GLZ2},
which is based on the non-Abelian symmetry group $O(2)$ \cite{su5},
and in the model of~\cite{GLD4},
which is based on the discrete symmetry $D_4$,
one finds $\theta_{13} = 0$ and $\theta_{23} = \pi / 4$;
the model of~\cite{kubo},
which is based on the permutation group $S_3$,
displays only a minimal deviation from those predictions.
\item The authors of~\cite{A4},
on the basis of a symmetry $A_4$,
found $\theta_{23} = \pi / 4$
but their $\theta_{13}$
is non-zero and can even be rather large.
\item With the non-standard CP transformation used in~\cite{CP}
atmospheric mixing is maximal but $\theta_{13}$ is free.
\item Conversely,
it was shown in~\cite{low} that one can construct models
based on Abelian groups,
in particular on $\mathbbm{Z}_4$,
which display $\theta_{13} = 0$  but have a free atmospheric mixing angle.
\end{itemize}
The listing above demonstrates that,
without specifying the family symmetry group
and other details of the model,
no general statement can be made
as regards a possible relation between $\theta_{13}$ and $\theta_{23}$.

In this paper we want to discuss
a modification of the $D_4$ model of~\cite{GLD4}.
In this modification one breaks the $D_4$ symmetry softly
in such a way that $\theta_{13} = 0$ is preserved
but the atmospheric mixing angle becomes \emph{free}
and controlled by the strength of the soft symmetry-breaking term.
The model discussed in this paper has the same mixing features
as the models in~\cite{low},
but a completely different structure
due to the \emph{non}-Abelian symmetry group. 
Both the original and the softly broken $D_4$ models
belong to a class of models
where the Yukawa-coupling matrices are diagonal,
and lepton mixing originates exclusively
in the Majorana mass matrix $M_R$ of the heavy neutrinos;
the \emph{suppression of neutral flavour-changing interactions}
is a general feature of that class of models \cite{GLnondec}.
We furthermore show that,
in the softly broken $D_4$ model,
successful leptogenesis
can be traced back to the original $D_4$ model.

This paper is organized as follows.
In Section~\ref{D4model} we review the original $D_4$ model.
The softly broken $D_4$ model is introduced in Section~\ref{softD4},
where we also investigate its predictions for lepton mixing.
The neutrino mass spectrum,
the effective mass in neutrinoless $\beta\beta$ decay,
and leptogenesis are studied in Section~\ref{numasses}. 
In Section~\ref{radiative} we perform a qualitative discussion
of the renormalization-group corrections
to the prediction $\theta_{13} = 0$.
Our conclusions are found in Section~\ref{concl}.

\section{The original $D_4$ model} \label{D4model}

The $D_4$ model \cite{GLD4} has three lepton families, 
including three right-handed neutrinos
which enable the seesaw mechanism \cite{seesaw}.
We generically denote $e$,
$\mu$,
and $\tau$ by $\alpha$;
we have then three left-handed lepton doublets $D_{\alpha L}$,
three right-handed charged-lepton singlets $\alpha_R$,
and three right-handed neutrino singlets $\nu_{\alpha R}$.
The scalar sector consists of three Higgs doublets $\phi_1$,
$\phi_2$,
and $\phi_3$,
together with two \emph{real} gauge singlets $\chi_1$ and $\chi_2$.

The $D_4$ model has three symmetries of the $\mathbbm{Z}_2$ type:
\be
\begin{array}{rl}
\mathbbm{Z}_2^{(\tau)}: &
D_{\tau L},\ \tau_R,\ \nu_{\tau R},\ \chi_2\, \
\mathrm{change\ sign;}
\\
\mathbbm{Z}_2^{(\mathrm{tr})}: &
D_{\mu L} \leftrightarrow D_{\tau L},\
\mu_R \leftrightarrow \tau_R,\
\nu_{\mu R} \leftrightarrow \nu_{\tau R},\
\chi_1 \leftrightarrow \chi_2,\ 
\phi_3 \to - \phi_3; 
\\
\mathbbm{Z}_2^{(\mathrm{aux})}: &
\nu_{eR},\ \nu_{\mu R},\ \nu_{\tau R},\ \phi_1,\ e_R\, \
\mathrm{change\ sign.}
\end{array}
\label{symm}
\ee
The symmetry $\mathbbm{Z}_2^{(\tau)}$
flips the sign of all multiplets of the $\tau$ family,
while $\mathbbm{Z}_2^{(\mathrm{tr})}$
exchanges the multiplets of the $\mu$ and $\tau$ families. 
The auxiliary symmetry $\mathbbm{Z}_2^{(\mathrm{aux})}$ 
ensures that $\phi_2$ and $\phi_3$ do not have Yukawa couplings
to the $\nu_{\alpha R}$,
so that the spontaneous breaking of $\mathbbm{Z}_2^{(\mathrm{tr})}$,
which is necessary for $m_\mu \neq m_\tau$,
does not have consequences in the neutrino Dirac mass matrix $M_D$.

As discussed in~\cite{GLD4},
$\mathbbm{Z}_2^{(\tau)}$ and $\mathbbm{Z}_2^{(\mathrm{tr})}$
do not commute,
and together they generate a non-Abelian group $D_4$
with eight elements.
This group has five inequivalent irreducible representations (irreps):
one two-dimensional irrep and four one-dimensional irreps. 
It is clear that $\left( D_{\mu L}, D_{\tau L} \right)$,
$\left( \mu_R, \tau_R \right)$,
$\left( \nu_{\mu R}, \nu_{\tau R} \right)$,
and $\left( \chi_1, \chi_2 \right)$ transform as doublets of $D_4$,
whereas $\phi_1$ and $\phi_2$
transform according to the trivial one-dimensional irrep,
and $\phi_3$ according to a non-trivial one-dimensional irrep. 

The above multiplets and symmetries determine the Yukawa Lagrangian
\ba
\mathcal{L}_\mathrm{Y} &=&
- \left[ y_1 \bar D_{eL} \nu_{eR} + y_2 \left( \bar D_{\mu L} \nu_{\mu R} + 
\bar D_{\tau L} \nu_{\tau R} \right) \right]
\tilde \phi_1
\no & &
- y_3 \bar D_{eL} e_R \phi_1
- y_4 \left( \bar D_{\mu L} \mu_R + \bar D_{\tau L} \tau_R \right) \phi_2
- y_5 \left( \bar D_{\mu L} \mu_R - \bar D_{\tau L} \tau_R \right) \phi_3
\no & &
+ \frac{1}{2}\,y_\chi \nu_{eR}^T C^{-1} 
\left( \nu_{\mu R} \chi_1 + \nu_{\tau R} \chi_2 \right)
+ \mbox{H.c.},
\label{Y}
\ea
where $\tilde \phi_1 \equiv i \tau_2 \phi_1^\ast$;
they also determine the Majorana mass terms
\be\label{majorana}
\mathcal{L}_\mathrm{M} = \frac{1}{2} \left[
M^\ast \nu_{eR}^T C^{-1} \nu_{eR} +
{M^\prime}^\ast \left( \nu_{\mu R}^T C^{-1} \nu_{\mu R}
+ \nu_{\tau R}^T C^{-1} \nu_{\tau R} \right) \right] + \mbox{H.c.}
\ee
The vacuum expectation values (VEVs) 
$\left\langle 0 \left| \phi_j^0 \right| 0 \right\rangle
= v_j \left/ \sqrt{2} \right.$
($j = 1, 2, 3$)
fix the charged-lepton masses as
\ba
\sqrt{2} m_e &=& \left| y_3 v_1 \right|,
\no
\sqrt{2} m_\mu  &=& \left| y_4 v_2 + y_5 v_3 \right|,
\label{masses}
\\
\sqrt{2} m_\tau &=& \left| y_4 v_2 - y_5 v_3 \right|.
\ea
The neutrino Dirac mass matrix is given by
\be
M_D = {\rm diag} \left( a, b, b \right),
\label{MD}
\ee
with $\sqrt{2} a = y_1^\ast v_1$ and $\sqrt{2} b = y_2^\ast v_1$.
The charged-lepton mass matrix and $M_D$
are \emph{simultaneously} diagonal.
Therefore,
lepton mixing must result exclusively from a non-diagonal $M_R$,
and this arises by virtue of the non-zero VEVs of $\chi_{1,2}$.
Let us write those VEVs as
\be
\begin{array}{rcl}
\left\langle 0 \left| \chi_1 \right| 0 \right\rangle
&=& W \cos{\gamma},
\\*[1mm]
\left\langle 0 \left| \chi_2 \right| 0 \right\rangle &=& W \sin{\gamma},
\end{array}
\label{gamma}
\ee
with $W$ real and positive.
Then the Majorana mass matrix
for the right-handed neutrino singlets is
\be
M_R \left( \gamma \right)
= \left( \begin{array}{ccc}
M & M_\chi \cos{\gamma} & M_\chi \sin{\gamma} \\
M_\chi \cos{\gamma} & M' & 0 \\
M_\chi \sin{\gamma} & 0 & M'
\end{array} \right),
\label{MRgamma}
\ee
where $M_\chi = y_\chi^\ast W$.
The effective Majorana mass matrix for the light neutrinos
is given by
\be
\mathcal{M}_\nu \left( \gamma \right)
= - M_D^T\, M_R \left( \gamma \right)^{-1} \! M_D,
\label{Mnu}
\ee
where $M_D = M_D^T$ is in equation~(\ref{MD}).
Since the charged-lepton mass matrix is diagonal,
the lepton mixing matrix $U$ is found simply
through the diagonalization procedure 
\be
U^T \mathcal{M}_\nu \left( \gamma \right) U = 
\mbox{diag} \left( m_1, m_2, m_3 \right),
\ee
with real and non-negative masses $m_j$
($j = 1, 2, 3$)
for the light neutrinos.

In the $D_4$-invariant scalar potential $V_\mathrm{sym}$, 
the $\chi$-dependent terms are \cite{GLD4}
\ba
V_\mathrm{sym} &=& \cdots + \left( \chi_1^2 + \chi_2^2 \right)
\left[ - \mu + \rho_1 \phi_1^\dagger \phi_1
+ \rho_2 \left( \phi_2^\dagger \phi_2 + \phi_3^\dagger \phi_3
\right) \right] + \lambda \left( \chi_1^2 + \chi_2^2 \right)^2
\no & &
+ \left( \chi_1^2- \chi_2^2 \right)
\left( \eta \phi_2^\dagger \phi_3 + \eta^\ast \phi_3^\dagger \phi_2 \right)
+ \lambda^\prime \left( \chi_1^2- \chi_2^2 \right)^2,
\label{pot}
\ea
where $\mu$, 
$\rho_1$,
$\rho_2$,
$\lambda$,
$\eta$,
and $\lambda^\prime$ are c-numbers.
Then,
$\gamma$ is determined by the minimization of
\be
f_\mathrm{sym} \left( \gamma \right) =
\lambda^\prime W^4 \cos^2{2 \gamma} 
+ \mathrm{Re} \left( \eta v_2^\ast v_3 \right) W^2 \cos{2 \gamma}.
\ee
Provided $\lambda^\prime > 0$
and $\left| \mathrm{Re} \left( \eta v_2^\ast v_3 \right) \right|
\le 2 \lambda^\prime W^2$,
the minimum of $f$ is at
\be
\cos{2 \gamma} =
- \frac{\mathrm{Re} \left( \eta v_2^\ast v_3 \right)}
{2 \lambda^\prime W^2}.
\label{gamma1}
\ee
According to the seesaw mechanism \cite{seesaw},
we assume
\be
\left| M \right|, \left| M^\prime \right| \gg v,
\label{scales}
\ee
where $v = \sqrt{|v_1|^2 + |v_2|^2 + |v_3|^2} \simeq 246\, \mathrm{GeV}$.
We furthermore assume $W \sim \left| M \right|, \left| M^\prime \right|$,
so that the off-diagonal matrix elements
of $M_R \left( \gamma \right)$ are not much smaller
than the diagonal matrix elements,
lest the solar mixing angle becomes very small.
Under these assumptions,
$\left| \cos{2 \gamma} \right|$ is negligibly small and, 
for all practical purposes, 
\be
\left\langle 0 \left| \chi_1 \right| 0 \right\rangle =
\left\langle 0 \left| \chi_2 \right| 0 \right\rangle =
\frac{W}{\sqrt{2}},
\label{VEVchi}
\ee
or $\gamma = \pi / 4$.
Corrections to this value of $\gamma$ are of order $v^2 / W^2$.

One can easily check that \cite{ustron03}
\be
S M_R \left( \pi / 4 \right) S = M_R \left( \pi / 4 \right),
\ee
where
\be
S = \left( \begin{array}{ccc} 1 & 0 & 0 \\ 0 & 0 & 1 \\ 0 & 1 & 0
\end{array} \right).
\label{S}
\ee
Then,
due to equations~(\ref{Mnu}) and~(\ref{MD}),
also $S \mathcal{M}_\nu \left( \pi / 4 \right) S
= \mathcal{M}_\nu \left( \pi / 4 \right)$.
From this the predictions of the $D_4$ model follow \cite{GLD4}:
maximal atmospheric mixing angle,
i.e.\ $\theta_{23} = \pi / 4$;
free,
in general large,
solar mixing angle $\theta_{12}$;
and $\theta_{13} = 0$.
One also finds---see Section~\ref{numasses}---that,
because the $\left( \mu, \tau \right)$ matrix element
of $M_R \left( \gamma \right)$ vanishes,
the inverted neutrino mass spectrum is excluded.

\section{The softly broken $D_4$ model} \label{softD4}

In the original $D_4$ model of the previous section,
all three $\mathbbm{Z}_2$ symmetries of equation~(\ref{symm})
are spontaneously broken,
and none of them is broken in the Lagrangian itself.
In this section we introduce a soft dimension-two breaking
of $\mathbbm{Z}_2^{(\mathrm{tr})}$. 
It is easily found that there are two terms performing that breaking:
one of them is $\phi_2^\dagger \phi_3$
plus its Hermitian conjugate,
which is largely irrelevant;
the other one is
\be
V_\mathrm{soft} =
\mu_\mathrm{soft} \left( \chi_1^2 - \chi_2^2 \right),
\label{Vsoft}
\ee
which is to be added to the scalar potential.
Now the angle $\gamma$ of equation~(\ref{gamma})
is determined by the minimization of 
\be
f \left( \gamma \right) = f_\mathrm{sym} \left( \gamma \right)
+ \mu_\mathrm{soft} W^2 \cos{2 \gamma},
\ee
leading to 
\be
\cos{2 \gamma} =
- \frac{\mu_\mathrm{soft} + \mathrm{Re} \left( \eta v_2^\ast v_3 \right)}
{2 \lambda^\prime W^2}.
\ee
We assume $\left| \mu_\mathrm{soft} \right|$
to be of the same order of magnitude as $W$,
i.e.\ we assume it to be of the seesaw scale;
this results in a non-negligible deviation of $\gamma$ from $\pi / 4$.
As discussed in the previous section, 
such a deviation is in general present,
cf.\ equation~(\ref{gamma1}),
yet without the soft breaking in equation~(\ref{Vsoft})
it is very small because $v \ll W$.\footnote{We might as well assume $W$
to be of the Fermi scale,
and then $\mu_\mathrm{soft}$ would be unnecessary.
But this would imply the off-diagonal elements of $M_R$
being much smaller than the diagonal ones,
and this would prevent the fitting
of the observed large solar mixing angle.}

Defining 
\be
S \left( \gamma \right) = \left( \begin{array}{ccc}
1 & 0 & 0 \\ 0 & \cos{2 \gamma} & \sin{2 \gamma} \\ 
0 & \sin{2 \gamma} & - \cos{2 \gamma} \end{array} \right)
= S \left( \gamma \right)^{-1},
\label{Sgamma}
\ee
which generalizes the matrix $S$ in equation~(\ref{S}),
it is easy to check that 
\be
S \left( \gamma \right) M_R \left( \gamma \right)
S \left( \gamma \right) = M_R \left( \gamma \right) 
\ \Rightarrow \
S \left( \gamma \right) \mathcal{M}_\nu \left( \gamma \right)
S \left( \gamma \right) = \mathcal{M}_\nu \left( \gamma \right).
\label{invariance}
\ee
The orthogonal matrix $S \left( \gamma \right)$
has a unique eigenvalue $-1$,
corresponding to the eigenvector
\be
u_3 = \left( \begin{array}{c} 0 \\ - \sin{\gamma} \\
\cos{\gamma} \end{array} \right).
\label{u3}
\ee
Equation~(\ref{invariance}) then implies that $u_3$
is also an eigenvector of $\mathcal{M}_\nu \left( \gamma \right)$.
Since $u_3$ is real and its first entry is zero,
this eigenvector---possibly multiplied by a phase---must
constitute the third column of the lepton mixing matrix $U$. 
Explicit calculation of $\mathcal{M}_\nu \left( \gamma \right)$ using
\ba
M_R \left( \gamma \right)^{-1} &=&
\frac{1}{M^\prime \left( M M^\prime - M_\chi^2 \right)}
\no & & \times \left( \begin{array}{ccc}
{M^\prime}^2 &
- M^\prime M_\chi \cos{\gamma} &
- M^\prime M_\chi \sin{\gamma} \\
- M^\prime M_\chi \cos{\gamma} &
M M^\prime - M_\chi^2 \sin^2{\gamma} &
M_\chi^2 \sin{\gamma} \cos{\gamma} \\
- M^\prime M_\chi \sin{\gamma} &
M_\chi^2 \sin{\gamma} \cos{\gamma} &
M M^\prime - M_\chi^2 \cos^2{\gamma}
\end{array} \right)
\label{MR-1}
\ea
shows that the eigenvalue is $- b^2 / M^\prime$,
hence the mass of the third light neutrino is
\be
m_3 = \left| \frac{b^2}{M^\prime} \right|.
\ee

Thus,
\be
U_{e3} = 0 
\label{Ue3=0}
\ee
is \emph{exact} at the tree level,
while
\be
\sin^2{2\theta_\mathrm{atm}} = 
4 \left| U_{\mu 3} \right|^2
\left( 1 - \left| U_{\mu 3} \right|^2 \right)
= \sin^2{2 \gamma},
\ee
i.e.\ $\theta_\mathrm{atm} = \gamma$.
The atmospheric mixing angle is equal to the angle $\gamma$
from the VEVs of $\chi_{1,2}$ in equations~(\ref{gamma}).

We thus have a \emph{model}
where the atmospheric mixing angle is arbitrary, 
since its deviation from $\pi / 4$ is controlled
by the strength $\mu_\mathrm{soft}$
of the soft-breaking term $V_\mathrm{soft}$
in equation~(\ref{Vsoft});
on the other hand---at least
at the tree level---the angle $\theta_{13}$ vanishes.
The solar mixing angle is completely arbitrary,
but in general it will be large.

We have broken the $\mu$--$\tau$ interchange symmetry
$\mathbbm{Z}_2^{(\mathrm{tr})}$ softly in the scalar potential
by means of the term in equation~(\ref{Vsoft}).
That soft-breaking term is of dimension two.
One might,
in addition,
also break $\mathbbm{Z}_2^{(\mathrm{tr})}$
through terms of dimension three,
viz.\ by assuming $\left( M_R \right)_{\mu \mu}$ to be different
from $\left( M_R \right)_{\tau \tau}$.
A general mixing matrix $U$ would then follow.
It is,
however,
consistent to avoid soft-breaking terms of dimension three
while allowing those of dimension two.

\section{Neutrino masses and leptogenesis} \label{numasses}

The diagonalization of the mass matrices 
in equations~(\ref{MRgamma}) and~(\ref{Mnu})
can be related to the diagonalization of those matrices
when $\gamma = \pi / 4$.
Let us define the orthogonal matrix
\be
O = \left( \begin{array}{ccc}
1 & 0 & 0 \\ 
0 & \left( \cos{\gamma} + \sin{\gamma} \right) \left/ \sqrt{2} \right.
& \left( \cos{\gamma} - \sin{\gamma} \right) \left/ \sqrt{2} \right. \\ 
0 & \left( \sin{\gamma} - \cos{\gamma} \right) \left/ \sqrt{2} \right.
& \left( \cos{\gamma} + \sin{\gamma} \right) \left/ \sqrt{2} \right.
\end{array} \right).
\label{O}
\ee
Then we easily find that
\be\label{sigma}
O^T M_R \left( \gamma \right) O = M_R \left( \pi / 4 \right)
\ \Rightarrow \
O^T \mathcal{M}_\nu \left( \gamma \right) O
= \mathcal{M}_\nu \left( \pi / 4 \right).
\ee
The matrix
\be
U^\prime = e^{i \hat \alpha}
\left( \begin{array}{ccc}
c_{12} & s_{12} & 0 \\
- s_{12} / \sqrt{2} & c_{12} / \sqrt{2} & - 1 / \sqrt{2} \\
- s_{12} / \sqrt{2} & c_{12} / \sqrt{2} & 1 / \sqrt{2}
\end{array} \right) e^{i \hat \beta},
\label{U'}
\ee
where
\be
e^{i \hat \alpha} = \mbox{diag}
\left( 1, e^{i\alpha}, e^{i\alpha} \right),
\label{alpha}
\ee
$c_{12} \equiv \cos{\theta_{12}}$,
and $s_{12} \equiv \sin{\theta_{12}}$,
diagonalizes $\mathcal{M}_\nu \left( \pi / 4 \right)$ \cite{GLD4}.
The matrices $e^{i \hat \alpha}$ and $e^{i \hat \beta}$
are diagonal phase matrices;
the first of them is unphysical,
and it can be shown that it is of the form given
in equation~(\ref{alpha}).
According to equation~(\ref{sigma}),
the matrix $U$ which diagonalizes
$\mathcal{M}_\nu \left( \gamma \right)$ is
\be
U = O\, U^\prime = e^{i \hat \alpha}
\left( \begin{array}{ccc}
c_{12} & s_{12} & 0 \\
- s_{12} \cos{\gamma} & c_{12} \cos{\gamma} & - \sin{\gamma} \\
- s_{12} \sin{\gamma} & c_{12} \sin{\gamma} & \cos{\gamma}
\end{array} \right)
e^{i \hat \beta},
\label{U}
\ee
cf.\ equation~(\ref{u3}).
With $e^{i \hat \beta} = \mbox{diag} \left( e^{i \beta_1},\,
e^{i \beta_2},\, e^{i \beta_3} \right)$,
the physical Majorana phases
can be chosen to be $\Delta \equiv 2 \left( \beta_1 - \beta_2 \right)$
and $2 \left( \beta_1 - \beta_3 \right)$.

A salient feature of the $D_4$ model---whether
softly broken or not---is the fact that
$\left( M_R \right)_{\mu \tau} = 0$.
This translates into \cite{GLD4}
\ba
0 &=& \sum_{j = 1}^3 m_j^{-1} U_{\mu j} U_{\tau j}
\no &=& 
e^{2 i \alpha} \sin{\gamma} \cos{\gamma} 
\left( \frac{s_{12}^2 e^{2i\beta_1}}{m_1} +
\frac{c_{12}^2 e^{2i\beta_2}}{m_2} -
\frac{e^{2i\beta_3}}{m_3} \right) = 0.
\label{cond}
\ea
We see that the expression within parentheses is zero
irrespective of the value of $\gamma$.
Hence,
the effective mass probed in neutrinoless $\beta\beta$ decay,
$\left| \left\langle m \right\rangle \right|
= \left| \left( \mathcal{M}_\nu \right)_{ee} \right|$,
is,
just as in~\cite{GLD4},
\be
\left| \left\langle m \right\rangle \right| = 
\left| m_1 c^2_{12} e^{- 2 i \beta_1} +
m_2 s^2_{12} e^{- 2 i \beta_2} \right| =
\frac{m_1 m_2}{m_3}.
\label{meff}
\ee
Moreover,
just as in the original $D_4$ model, 
only the normal spectrum $m_1 < m_2 < m_3$ is allowed.
($m_1 < m_2$ holds by definition,
and the solar mass-squared difference
is $\Delta m^2_\odot = m_2^2 - m_1^2$.)

For baryogenesis via leptogenesis
\cite{yanagida}---for reviews,
see for instance~\cite{lepto-reviews}---we must
diagonalize the mass matrix $M_R$.
Denoting the unitary diagonalization matrix by $V$
and the heavy-neutrino masses by $M_j$,
one has
\be
V^T M_R \left( \gamma \right) V
= {V^\prime}^T M_R \left( \pi / 4 \right) V^\prime
= \mbox{diag} \left( M_1,\, M_2,\, M_3 \right),
\ee
with $V = O\, V^\prime$.
The matrix which enters the calculation of the CP asymmetry
generated by the decay of the heavy neutrinos
is given by \cite{lepto-reviews}
\be
R = V^T \! M_D M_D^\dagger V^\ast
= {V^\prime}^T \! M_D M_D^\dagger {V^\prime}^\ast.
\label{uipto}
\ee
The second equality in equation~(\ref{uipto})
is justified by the twofold degeneracy of $M_D$,
see equation~(\ref{MD}).
Thus,
in the softly broken $D_4$ model only $V^\prime$ is relevant
for leptogenesis.
As emphasized before,
$U^\prime$ and $V^\prime$ coincide
with the diagonalization matrices of the original $D_4$ model.
Hence,
the soft breaking of $D_4$ \emph{does not alter}
the expression of the lepton asymmetry derived in~\cite{GLleptogenesis},
and one obtains the same results
irrespective of the value of $\gamma = \theta_{23}$.
The masses of the third-generation neutrinos,
both in the light- and heavy-neutrino sectors
(i.e., both $m_3$ and $M_3$),
do not occur in the expression for leptogenesis---which depends only
on $m_{1,2}$,
$M_{1,2}$,
$\theta_{12}$,
and $\Delta$ \cite{GLleptogenesis,NOON04}.
However,
$m_3$ enters leptogenesis indirectly,
because the Majorana phase $\Delta$ may be expressed
as a function of
$m_3$; indeed,
from equation~(\ref{cond}) \cite{GLleptogenesis},
\be
\left| m_2 s_{12}^2 e^{i \Delta} + m_1 c_{12}^2 \right|
= \frac{m_1 m_2}{m_3},
\ee
hence
\be
\cos \Delta = 
\frac{\left( m_1 m_2 / m_3 \right)^2
- c_{12}^4 m_1^2 - s_{12}^4 m_2^2}{2 c_{12}^2 s_{12}^2 m_1 m_2}.
\label{Delta}
\ee
Successful leptogenesis in the $D_4$ model requires \cite{GLleptogenesis}
the lightest-neutrino mass $m_1$ to be below $10^{-2}\, \mathrm{eV}$.
In that region equation~(\ref{Delta})
gives a strong restriction on the allowed range for $m_1$,
since $\cos \Delta$ must be larger than $-1$;
using $\theta_{12} = 33^\circ$,
$\Delta m^2_\odot = 7.1 \times 10^{-5}\, \mathrm{eV}^2$,
and $\Delta m^2_\mathrm{atm} = m_3^2 - m_1^2
= 2 \times 10^{-3}\, \mathrm{eV}^2$ \cite{maltoni},
one obtains $2.9 \times 10^{-3}\, \mathrm{eV}
\, \lesssim\,  m_1
\, \lesssim\,  7.1 \times 10^{-3}\, \mathrm{eV}$.
In order to reproduce the baryon-over-photon ratio $\eta_B$,
$M_1$ must lie in between $10^{11}$ and $10^{12}\, \mathrm{GeV}$,
if we assume $M_2 \gg M_1$.
The maximum value of $\eta_B$ is attained
for $m_1 \simeq 4 \times 10^{-3}\, \mathrm{eV}$,
and the experimental value 
$\eta_B \sim 6.5 \times 10^{10}$ \cite{spergel}
can easily be reproduced---see figure~1 of~\cite{NOON04}.

\section{Radiative corrections} \label{radiative}

The form of $\mathcal{M}_\nu$ given by equations~(\ref{MD}),
(\ref{MRgamma}),
and (\ref{Mnu}) holds only at the seesaw scale $m_R$.
Radiative corrections must be taken into account
if one wants to calculate $\mathcal{M}_\nu$
at the electroweak scale.
The effective operators relevant in this context
are~\cite{babu,antusch1,antusch2}
\ba
\mathcal{O}_{ij} &=&
\sum_{\alpha,\beta} \sum_{a,b,c,d}
\left[ \left( D_{\alpha L} \right)^T_a
\kappa^{(ij)}_{\alpha \beta}
C^{-1} \left( D_{\beta L} \right)_c \right]
\left[ \varepsilon^{ab} \left( \phi_i \right)_b \right]
\left[ \varepsilon^{cd} \left( \phi_j \right)_d \right]
\label{uryte} \\ &=& \sum_{\alpha,\beta}
\kappa_{\alpha \beta}^{(ij)}
\left( \nu_{\alpha L}^T \phi_i^0 - \alpha_L^T \phi_i^+ \right)
C^{-1} \left( \nu_{\beta L} \phi_j^0 - \beta_L \phi_j^+ \right),
\label{Op}
\ea
where $a,b,c,d$ are $SU(2)$ indices,
$\varepsilon = i \tau_2$ is the $2 \times 2$ antisymmetric tensor,
and the $\kappa^{(ij)}$ are matrices in family
space.
Note that we may,
without loss of generality,
enforce the condition
$\mathcal{\kappa}^{(ij)}_{\alpha \beta}
= \mathcal{\kappa}^{(ji)}_{\beta \alpha}$.
We want to discuss the renormalization-group (RG)
evolution of the effective coupling matrices $\kappa^{(ij)}$
from the seesaw scale $m_R$ down to the electroweak scale;
the latter may be represented by the $Z^0$ mass $m_Z$. 
Our aim is to estimate a possible modification
of equation~(\ref{Ue3=0}) by the RG evolution.
We denote the dimensionless variable of the RG by $t$;
the RG evolution goes from $t_0 = \ln{\left( m_R / m_Z \right)}$
to $t_1 = 0$.
The initial conditions for the RG equations are
\be
\begin{array}{rcl}
\kappa^{(11)} \left( t_0 \right) &=& {\displaystyle \frac{1}{v_1^2}}\,
\mathcal{M}_\nu \left( \gamma \right),
\\*[4mm]
\kappa^{(ij)} \left( t_0 \right) &=& 0\
\mbox{for\ all\ other\ } (ij).
\end{array}
\label{initial}
\ee
In equation~(\ref{initial}) we have used the fact that
in our model only the Higgs doublet $\phi_1$ has Yukawa couplings
to the right-handed neutrinos,
and gives thereby rise to $M_D$.

The symmetry $\mathbbm{Z}_2^{(\mathrm{aux})}$,
which is preserved throughout the RG evolution
since it is only broken spontaneously at the electroweak scale,
implies that there is a symmetry $\phi_1 \to - \phi_1$
in the evolution of the operators in equation~(\ref{Op}).
Since at the high scale only $\mathcal{O}_{11}$ is present,
which is invariant under $\mathbbm{Z}_2^{(\mathrm{aux})}$,
it follows that the operators
$\mathcal{O}_{12}$ and $\mathcal{O}_{13}$,
which are odd under $\mathbbm{Z}_2^{(\mathrm{aux})}$,
remain zero at all scales.
At the electroweak scale we will then have
\be
\mathcal{M}_\nu = 
\sum_{i=1}^3 v_i^2 \kappa^{(ii)} \left( t_1 \right)
+ v_2 v_3 \left[ \kappa^{(23)} \left( t_1 \right)
+ \kappa^{(32)} \left( t_1 \right) \right].
\label{RGEMnu}
\ee

The matrix $\mathcal{M}_\nu$ at the electroweak scale
will not in general yield $U_{e3} = 0$.
Indeed,
using $U^T \! \mathcal{M}_\nu U = \mathrm{diag} \left(
m_1, m_2, m_3 \right)$,
we find that $U_{e3} = 0$ if and only if there is a vector $u$
with a zero first entry
such that $\mathcal{M}_\nu u = z u^\ast$,
where $z$ is a complex number and $m_3 = |z|$.
This means that the symmetric matrix $\mathcal{M}_\nu$ satisfies
\be
\left[ \left| \left( \mathcal{M}_\nu \right)_{e \mu} \right|^2
- \left| \left( \mathcal{M}_\nu \right)_{e \tau} \right|^2 \right]
\left( \mathcal{M}_\nu \right)_{\mu \tau}
=
\left( \mathcal{M}_\nu \right)_{e \mu}^\ast
\left( \mathcal{M}_\nu \right)_{e \tau}
\left( \mathcal{M}_\nu \right)_{\mu \mu}
-
\left( \mathcal{M}_\nu \right)_{e \mu}
\left( \mathcal{M}_\nu \right)_{e \tau}^\ast
\left( \mathcal{M}_\nu \right)_{\tau \tau}.
\ee
This condition will in general not be satisfied
by the matrix $\mathcal{M}_\nu$ at the electroweak scale
given in equation~(\ref{RGEMnu}),
but it \emph{is} satisfied by the matrix $\mathcal{M}_\nu$
at the seesaw scale given by equations~(\ref{MD}),
(\ref{MRgamma}),
and (\ref{Mnu}).

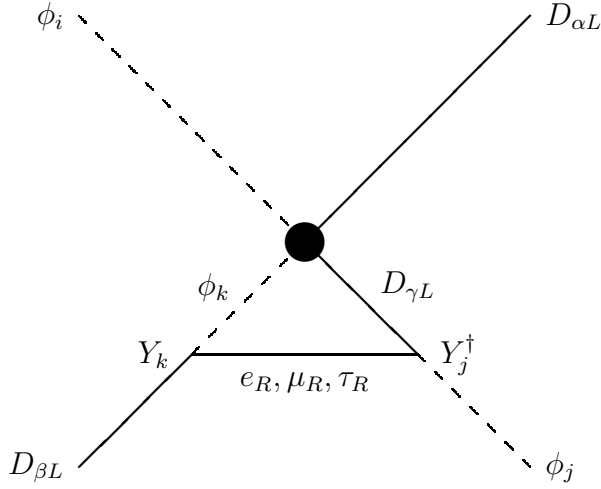
\begin{figure}
\setlength{\unitlength}{1cm}
\begin{center}
\begin{picture}(8,8)
\put(4,4){\circle*{5}}
\dashline{0.15}(4,4)(1,7)
\dashline{0.15}(5.5,2.5)(7,1)
\dashline{0.15}(4,4)(2.5,2.5)
\thicklines
\drawline(4,4)(5.5,2.5)
\drawline(4,4)(7,7)
\drawline(2.5,2.5)(1,1)
\drawline(2.5,2.5)(5.5,2.5)
\put(0.8,7){\makebox(0,0)[r]{$\phi_i$}}
\put(7.2,7){\makebox(0,0)[l]{$D_{\alpha L}$}}
\put(7.2,1){\makebox(0,0)[l]{$\phi_j$}}
\put(0.8,1){\makebox(0,0)[r]{$D_{\beta L}$}}
\put(4.0,2){\makebox(0,0)[b]{$e_R, \mu_R, \tau_R$}}
\put(5.0,3.2){\makebox(0,0)[lb]{$D_{\gamma L}$}}
\put(3.0,3.2){\makebox(0,0)[rb]{$\phi_k$}}
\put(2.2,2.5){\makebox(0,0)[r]{$Y_k$}}
\put(5.8,2.5){\makebox(0,0)[l]{$Y_j^\dagger$}}
\end{picture}
\end{center}
\caption{A typical vertex correction
which introduces flavour dependence
in the renormalization of the operators $\mathcal{O}_{ij}$
of equation~(\ref{Op}).
The relevant Yukawa-coupling matrices are indicated.
\label{fig}}
\end{figure}

Denoting the set of all matrices $\kappa^{(ij)}$ by $\kappa$, 
the differential equation for their RG evolution 
can symbolically be written as 
\be
16 \pi^2\,
\frac{\mathrm{d}\kappa (t)}{\mathrm{d}t} = 
L_t \left[ \kappa(t) \right],
\label{symbolical}
\ee
where $L_t$ is a linear operator acting on the $\kappa^{(ij)}$.
In $L_t$,
the Yukawa and gauge coupling constants appear in second order 
and the quartic Higgs couplings in first order.
The solution of equation~(\ref{symbolical})
is formally given by the series 
\be
\kappa(t) = \kappa(t_0) + \frac{1}{16 \pi^2} \int_{t_0}^t
\mathrm{d}t^\prime L_{t^\prime} \left[ \kappa(t_0) \right] + 
\left( \frac{1}{16 \pi^2} \right)^2 
\int_{t_0}^t \mathrm{d}t^\prime \int_{t_0}^{t^\prime} 
\mathrm{d}t^{\prime\prime}
L_{t^\prime} \left[
L_{t^{\prime\prime}} \left[ \kappa(t_0) \right] \right]
+ \cdots
\label{expansion}
\ee
Flavour dependence in ${\dot \kappa}^{(ij)}$
(the dot denotes the derivative relative to $t$)
can only be introduced by the Yukawa couplings.
One can easily convince oneself that only vertex corrections
of the type depicted in figure~\ref{fig}
need a closer look at.\footnote{All the types of graphs
relevant for the computation of the full RG equations for the
$\kappa^{(ij)}$ are depicted in~\cite{babu,antusch2}.}
The Yukawa-coupling matrices of the Higgs doublets $\phi_j$
to the right-handed charged leptons are given
by the second line of the Yukawa Lagrangian in equation~(\ref{Y});
they are
\ba
Y_1 &=& \mbox{diag} \left( y_3^\ast, 0, 0 \right), \no
Y_2 &=& \mbox{diag} \left( 0, y_4^\ast, y_4^\ast \right),
\label{Yuk} \\
Y_3 &=& \mbox{diag} \left( 0, y_5^\ast, - y_5^\ast \right).
\nonumber
\ea
One finds from figure~\ref{fig} that
${\dot \kappa}^{(ij)} + {\dot \kappa}^{(ji)}$ 
obtains a contribution
$\sum_k \left[ \kappa^{(ik)} Y_j^\dagger Y_k
+ \kappa^{(kj)} Y_i^\dagger Y_k \right]$ plus 
the transposed term.

A non-zero $\kappa^{(23)}$ is induced from $\kappa^{(11)}(t_0)$
in two steps.
Firstly,
vertex corrections from the quartic terms 
\be
\lambda^\prime \left( \phi_1^\dagger \phi_2 \right)^2 +
\lambda^{\prime\prime} \left( \phi_1^\dagger \phi_3 \right)^2
\ee
in the Higgs potential induce non-zero matrices $\kappa^{(22)}$ and
$\kappa^{(33)}$ via 
\ba
16 \pi^2\, \frac{\mathrm{d}\kappa^{(22)}}{\mathrm{d}t} &\sim& 
\lambda^\prime \kappa^{(11)}, 
\label{11} \\
16 \pi^2\, \frac{\mathrm{d}\kappa^{(33)}}{\mathrm{d}t} 
&\sim& \lambda^{\prime\prime} \kappa^{(11)}.
\label{22}
\ea
Secondly,
from a vertex correction of the type in figure~\ref{fig} one obtains 
\be\label{23}
16 \pi^2\, 
\frac{\mathrm{d}}{\mathrm{d}t} 
\left[ \kappa^{(23)} + \kappa^{(32)} \right]
\sim 
\kappa^{(22)} Y_3^\dagger Y_2 +
Y_2^T Y_3^\ast \kappa^{(22)} +
\kappa^{(33)} Y_2^\dagger Y_3 +
Y_3^T Y_2^\ast \kappa^{(33)}.
\ee

Equations~(\ref{11}),
(\ref{22}),
and~(\ref{23}) tell us
that, in the expansion of equation~(\ref{expansion}), the matrices
$\kappa^{(23)}$ and $\kappa^{(32)}$
appear first at second order.
Since
\be
Y_3^\dagger Y_2 = \left( Y_2^\dagger Y_3 \right)^\dagger = 
y_4^\ast y_5\, \mathrm{diag} \left( 0, +1, -1 \right),
\ee
the vector $u_3$ of equation~(\ref{u3}) will not be eigenvector of 
$\kappa^{(23)} + \kappa^{(32)}$.
One can check that,
at second order in the expansion of equation~(\ref{expansion}),
$u_3$ is still an eigenvector of the $\kappa^{(ii)}$ ($i=1,2,3$). 
Therefore,
we estimate that
the terms in $\mathcal{M}_\nu$ which are responsible for $U_{e3} \neq 0$
will typically be suppressed by a factor of the order of
\be
\frac{1}{(16 \pi^2)^2} 
\left( \ln \frac{m_R}{m_Z} \right)^2 
\left( \frac{m_\tau}{|v_2|} \right)^2 
\left| \lambda^{\prime,\prime\prime} \right|.
\label{estimate}
\ee
In this equation we have taken into account 
that there are two integrations over 
the interval of length $t_0$ and that the relevant 
Yukawa couplings are of the order $m_\tau/|v_2|$.
Numerically,
using $t_0 \sim 10$ and
even if we take the quartic Higgs couplings to be of order one, the
suppression factor is at least $10^{-4}$. Therefore, to a good approximation
we will still have $U_{e3} = 0$
at the electroweak scale.\footnote{The details of the RGE for
multiple-Higgs-doublet models will be published elsewhere.}

\section{Conclusions} \label{concl}

In this paper we have modified the $D_4$ model of~\cite{GLD4}
by allowing a soft breaking
of the subgroup $\mathbbm{Z}_2^{(\mathrm{tr})}$ of $D_4$,
i.e.\ by breaking the $\mu$--$\tau$ exchange symmetry softly---see
equation~(\ref{symm}).
An essential ingredient of the $D_4$ model
is the $D_4$ doublet $\left( \chi_1, \chi_2\right)$
of real scalar gauge singlets,
with VEVs and masses of the seesaw scale $m_R$.
It is the VEVs of those scalars which induce lepton mixing.
In particular,
one has
\be
\tan{\theta_{23}} = \frac{\langle \chi_2 \rangle_0}
{\langle \chi_1 \rangle_0}.
\ee
Requiring the soft breaking of $\mathbbm{Z}_2^{(\mathrm{tr})}$
to occur exclusively through terms of dimension two in the Lagrangian,
we find that the term of equation~(\ref{Vsoft})
appears in the scalar potential.
Whereas without soft breaking the VEVs of $\chi_1$ and $\chi_2$
are (almost) equal,
the soft breaking of $\mathbbm{Z}_2^{(\mathrm{tr})}$
induces a deviation from this equality.
Thus, the strength of the soft-breaking term in equation~(\ref{Vsoft})
determines the deviation of $\theta_{23}$ from $45^\circ$,
small deviations being natural in a technical sense.

We have demonstrated that the soft $D_4$ breaking
has no effect on the neutrino mass spectrum,
on the effective mass $| \langle m \rangle |$
in neutrinoless $\beta\beta$ decay,
and on leptogenesis.
Just as in the original $D_4$ model,
we predict a normal spectrum $m_1 < m_2 < m_3$, 
$\left| \langle m \rangle \right| < 10^{-2}\, \mathrm{eV}$,
and we find that leptogenesis is successful
when $m_1 \sim 4 \times 10^{-3}\, \mathrm{eV}$.

Like in all multi-Higgs models
where the Yukawa-coupling matrices are diagonal
and lepton mixing originates solely in $M_R$,
neutral flavour-changing interactions are suppressed \cite{GLnondec}:
on the one hand,
the amplitudes for e.g.\ $\mu^- \to e^- \gamma$ and $Z^0 \to e^+ \mu^-$
are proportional to $m_R^{-2}$,
hence highly suppressed;
on the other hand,
a decay like $\mu^- \to e^- e^+ e^-$
is suppressed by Yukawa couplings
but might fall in the discovery range of forthcoming experiments. 

The $D_4$ model---whether softly broken or not---predicts
$\theta_{13} = 0$.
We have estimated,
through renormalization-group methods,
the radiative corrections to this relation
to yield $\theta_{13} \sim 10^{-4}$
or smaller---the same estimate
also applies to the $\mathbbm{Z}_2$ model of \cite{GLZ2}.
The solar mixing angle is not predicted in the $D_4$ model;
without finetuning it will be large.

In summary,
we have constructed an extension of the Standard Model
based on the non-Abelian family symmetry group $D_4$.
Our model features a vanishing $\theta_{13}$
together with non-maximal atmospheric mixing.
The model is in agreement with all the existing data on neutrinos
and allows successful leptogenesis.

\vspace{6mm}

\noindent \textbf{Acknowledgements}:
The work of LL has been supported
by the Portuguese \textit{Funda\c c\~ao para a Ci\^encia e a Tecnologia}
under the project CFIF--Plurianual.



\begin{thebibliography}{99}

\bibitem{pontecorvo}
S.M. Bilenky and B. Pontecorvo, Phys. Rep. 41 (1978) 225; \\
S.M. Bilenky and S.T. Petcov, Rev. Mod. Phys. 59 (1987) 671.

\bibitem{MNS}
Z. Maki, M. Nakagawa, and S. Sakata, 
Prog. Theor. Phys. 28 (1962) 870.

\bibitem{maltoni}
M. Maltoni, T. Schwetz, M.A. T\'ortola, and J.W.F. Valle,
hep-ph/0405172,
to appear in the focus issue on Neutrino Physics
edited by F. Halzen, M. Lindner, and A. Suzuki,
New Journal of Physics.

\bibitem{altarelli}
G. Altarelli and F. Feruglio,
hep-ph/0405048, 
to appear in the focus issue on Neutrino Physics
edited by F. Halzen, M. Lindner, and A. Suzuki,
New Journal of Physics.

\bibitem{seesaw}
T. Yanagida,
in \textit{Proceedings of the workshop on unified theories
and baryon number in the universe (Tsukuba, Japan, 1979)},
eds. O. Sawada and A. Sugamoto
(Tsukuba: KEK report 79-18, 1979); \\ 
S.L. Glashow,
in \textit{Quarks and leptons,
proceedings of the advanced study institute (Carg\`ese, Corsica, 1979)},
eds. J-L. Basdevant et al.
(New York: Plenum, 1981); \\
M. Gell-Mann, P. Ramond, and R. Slansky,
in \textit{Supergravity},
eds. D.Z. Freedman and F. van Nieuwenhuizen
(Amsterdam: North Holland, 1979); \\
R.N. Mohapatra and G. Senjanovi\'c,
Phys. Rev. Lett. 44 (1980) 912.

\bibitem{GLZ2}
W. Grimus and L. Lavoura,
JHEP 07 (2001) 045 [hep-ph/0105212]; \\
W. Grimus and L. Lavoura,
Acta Phys. Polon. B 32 (2001) 3719 [hep-ph/0110041].

\bibitem{su5}
W. Grimus and L. Lavoura,
Eur. Phys. J. C 28 (2003) 123 [hep-ph/0211334].

\bibitem{GLD4}
W. Grimus and L. Lavoura, 
Phys. Lett. B 572 (2003) 189 [hep-ph/0305046].

\bibitem{kubo}
J. Kubo, 
Phys. Lett. B 578 (2004) 156 [hep-ph/0309167].

\bibitem{A4}
K.S. Babu, E. Ma, and J.W.F. Valle,
Phys. Lett. B 552 (2003) 207
[hep-ph/0206292].

\bibitem{CP}
W. Grimus and L. Lavoura,
Phys. Lett. B 579 (2004) 113 [hep-ph/0305309].

\bibitem{low}
C.I. Low, hep-ph/0404017.

\bibitem{GLnondec}
W. Grimus and L. Lavoura, 
Phys. Rev. D 66 (2002) 014016 [hep-ph/0204070].

\bibitem{ustron03}
W. Grimus and L. Lavoura,
Acta Phys. Polon. B 34 (2003) 5393 [hep-ph/0310050].

\bibitem{babu}
K.S. Babu, C.N. Leung, and J. Pantaleone, 
Phys. Lett. B 319 (1993) 191 [hep-ph/9309223].

\bibitem{antusch1}
S. Antusch, M. Drees, J. Kersten, M. Lindner, and M. Ratz,
Phys. Lett. B 519 (2001) 238 [hep-ph/0108005].

\bibitem{antusch2}
S. Antusch, M. Drees, J. Kersten, M. Lindner, and M. Ratz,
Phys. Lett. B 525 (2002) 130 [hep-ph/0110366].

\bibitem{yanagida}
M. Fukugita and T. Yanagida, 
Phys. Lett. B 174 (1986) 45.

\bibitem{lepto-reviews}
A. Pilaftsis, 
Int. J. Mod. Phys. A 14 (1999) 1811 [hep-ph/9812256]; \\
W. Buchm\"uller and M. Pl\"umacher, 
Int. J. Mod. Phys. A 15 (2000) 5047 
[hep-ph/0007176]; \\
E.A. Paschos, 
Pramana 62 (2004) 359 [hep-ph/0308261]; \\
W. Buchm\"uller, P. Di Bari, and M. Pl\"umacher,
hep-ph/0406014,
to appear in the focus issue on Neutrino Physics
edited by F. Halzen, M. Lindner, and A. Suzuki,
New Journal of Physics.

\bibitem{GLleptogenesis}
W. Grimus and L. Lavoura,
hep-ph/0311362,
to be published in J. Phys. G.

\bibitem{NOON04}
W. Grimus and L. Lavoura, 
hep-ph/0405261,
to be published in the Proceedings of NOON2004.

\bibitem{spergel}
D.N. Spergel et al., 
Astrophys. J. Suppl. 148 (2003) 175 [astro-ph/0302209].

\end{thebibliography}
\end{document}